\documentclass[aps,showpacs,twocolumn,superscriptaddress,prl,floatfix]{revtex4-1}
\usepackage{graphicx,color,epsfig,rotate}
\usepackage{amssymb,amsmath,bm,ulem}
\usepackage{dcolumn}  
\usepackage{hyperref} 
\usepackage[usenames,dvipsnames]{xcolor}
\usepackage{subfigure}
\usepackage{float}

\newcommand{\g} \textbf{}

\newcommand{\beginsupplement}{
        \setcounter{table}{0}
        \renewcommand{\thetable}{S\arabic{table}}
        \setcounter{figure}{0}
        \renewcommand{\thefigure}{S\arabic{figure}}}

\usepackage{bm}
\usepackage{graphicx}
\usepackage{color}
\usepackage{dcolumn}

\begin{document}
\title{Interaction Driven Subgap Spin Exciton in the Kondo Insulator SmB$_6$}  

\author{W. T. Fuhrman\normalfont\textsuperscript{$\star$}}
\email[]{wesleytf@gmail.com}
\affiliation{Institute for Quantum Matter and Department of Physics and Astronomy, The Johns Hopkins University, Baltimore, Maryland 21218 USA}

\author{J. Leiner\normalfont\textsuperscript{$\star$}}
\email[]{leinerjc@ornl.gov}
\affiliation{Quantum Condensed Matter Division, Oak Ridge National Laboratory, Oak Ridge, TN 37831}

\author{P. Nikoli\'{c}}
\affiliation{Institute for Quantum Matter and Department of Physics and Astronomy, The Johns Hopkins University, Baltimore, Maryland 21218 USA}
\affiliation{School of Physics, Astronomy and Computational Sciences,
George Mason University, Fairfax, VA 22030, USA}

\author{G. E. Granroth}
\affiliation{Neutron Data Analysis and Visualization Division, Oak Ridge National Laboratory, Oak Ridge, TN 37831}
\author{M. B. Stone }
\author{M. D. Lumsden}
\affiliation{Quantum Condensed Matter Division, Oak Ridge National Laboratory, Oak Ridge, TN 37831}

\author{L. DeBeer-Schmitt}
\affiliation{Instrument Source Division, Oak Ridge National Laboratory, Oak Ridge, TN 37831}

\author{P. A. Alekseev}
\affiliation{National Research Centre ``Kurchatov Institute", 123182 Moscow, Russia}
\affiliation{National Research Nuclear University "MEPhI", 115409 Moscow, Russia}

\author{J.-M. Mignot}
\affiliation{ Laboratoire L\'{e}on Brillouin, CEA-CNRS, CEA/Saclay, 91191 Gif sur Yvette, France}

\author{S. M. Koohpayeh}
\affiliation{Institute for Quantum Matter and Department of Physics and Astronomy, The Johns Hopkins University, Baltimore, Maryland 21218 USA}

\author{P. Cottingham}
\author{W. Adam Phelan}
\affiliation{Institute for Quantum Matter and Department of Physics and Astronomy, The Johns Hopkins University, Baltimore, Maryland 21218 USA}
\affiliation{Department of Chemistry, The Johns Hopkins University, Baltimore, Maryland 21218 USA}

\author{L. Schoop}
\affiliation{Institute for Quantum Matter and Department of Physics and Astronomy, The Johns Hopkins University, Baltimore, Maryland 21218 USA}
\affiliation{Department of Chemistry, Princeton University, Princeton, NJ 08540}

\author{T. M. McQueen}
\affiliation{Institute for Quantum Matter and Department of Physics and Astronomy, The Johns Hopkins University, Baltimore, Maryland 21218 USA}
\affiliation{Department of Chemistry, The Johns Hopkins University, Baltimore, Maryland 21218 USA}
\affiliation{Department of Materials Science and Engineering, The Johns Hopkins University, Baltimore, Maryland 21218 USA}

\author{C. Broholm}
\affiliation{Institute for Quantum Matter and Department of Physics and Astronomy, The Johns Hopkins University, Baltimore, Maryland 21218 USA}
\affiliation{Quantum Condensed Matter Division, Oak Ridge National Laboratory, Oak Ridge, TN 37831}
\affiliation{Department of Materials Science and Engineering, The Johns Hopkins University, Baltimore, Maryland 21218 USA}
\date{\today}

\begin{abstract}
Using inelastic neutron scattering, we map a 14 meV coherent resonant mode in the topological Kondo insulator SmB$_6$ and describe its relation to the low energy insulating band structure.  The resonant intensity is confined to the $X$ and $R$ high symmetry points, repeating outside the first Brillouin zone and dispersing less than 2 meV, with a 5$d$-like magnetic form factor.  We present a slave-boson treatment of the Anderson Hamiltonian with a third neighbor dominated hybridized band structure.  This approach produces a spin exciton below the charge gap with features that are consistent with the observed neutron scattering. We find that maxima in the wave vector dependence of the inelastic neutron scattering indicate band inversion. 
\end{abstract}

\maketitle
Recent theoretical work suggests SmB$_6$ could be a topological Kondo insulator (TKI), with an insulating bulk at low temperatures and a topologically protected metallic surface  \cite{TKI_theory1, takimoto2011smb6, PhysRevB.85.045130, TKI_theory_2,  theory_7_13, CubicTKI, legner2014topological} that was previously ascribed to impurities \cite{EarlyMetal2}. Because strong electron-electron interactions produce the insulating state, the surface may support exotic correlated physics \cite{roy2014surface, nikolic2014two, fuhrman2014gap}.  

Experimental investigations \cite{resistance1, kim2013surface, li2013quantum, neupane2013surface, jiang2013observation, denlinger2013temperature, adamPRX}, particularly spin-resolved angle-resolved photo-emission spectroscopy (ARPES) \cite{xu2014direct}, have provided compelling evidence that SmB$_6$ is a TKI.   However, information about the band structure within $\approx$ 50 meV of  the Fermi level is limited due to the polar surface, multiplet structure, and strong correlations. In this energy range the magnetic neutron scattering is sensitive to the renormalized band structure through the imaginary part of the momentum (${\bf Q}$) and energy ($\hbar\omega$) dependent generalized susceptibility. 

In this Letter, we present a comprehensive measurement of the inelastic magnetic neutron scattering cross section covering the full Brillouin zone of SmB$_6$ for energies below 50 meV.  We pair our experimental results with a slave-boson treatment of an Anderson Hamiltonian, and discuss how pseudonesting conditions for the renormalized band structure can be examined to corroborate a topologically nontrivial band structure for SmB$_6$. 

\begin{figure*}
\includegraphics[totalheight=0.385\textheight,viewport= 0 610 1650 1435, clip]{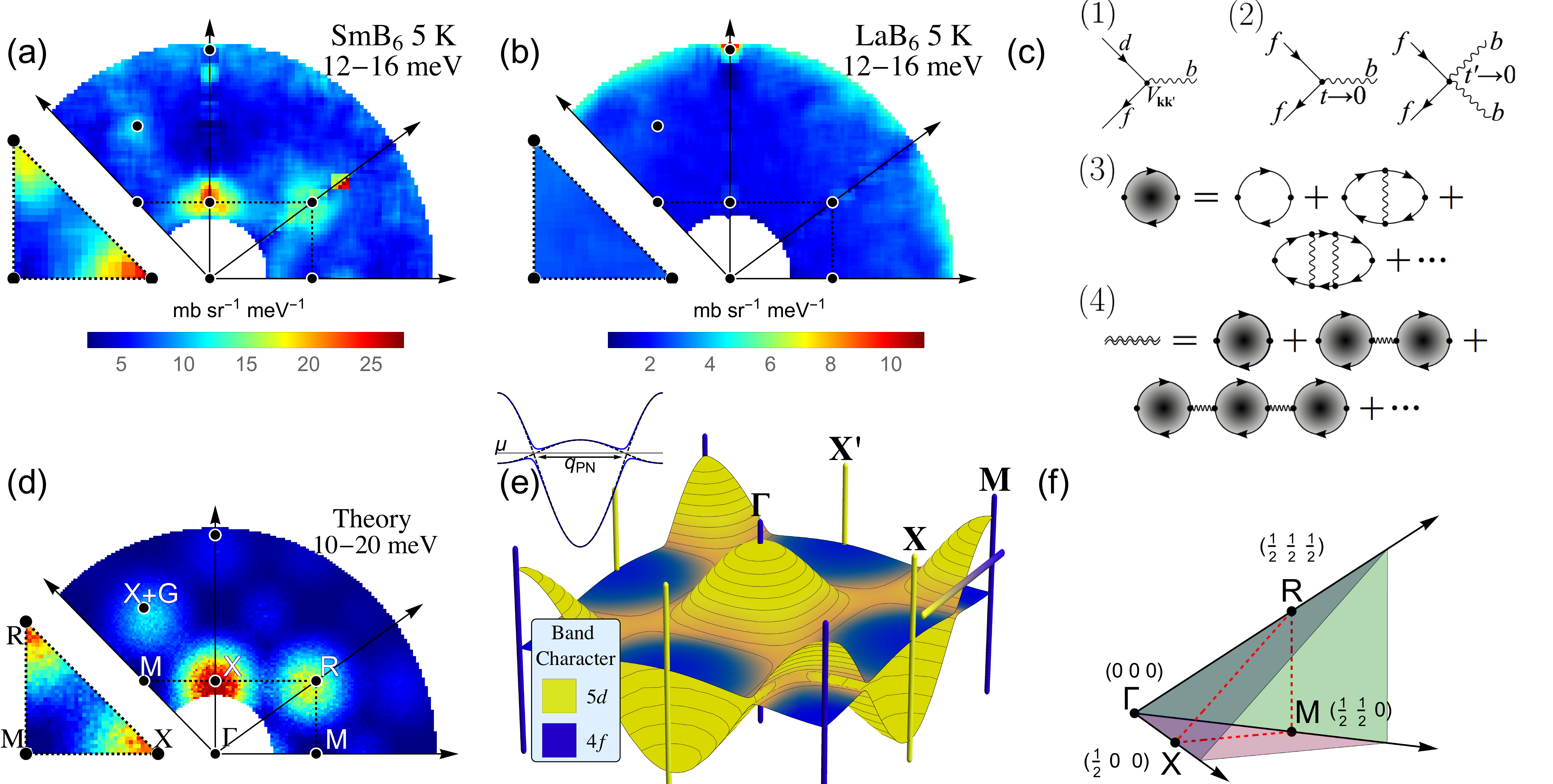}
\caption{ \label{14 meV} Energy integrated neutron scattering intensity (a, b, d) in high symmetry planes.  (a) $^{154}$Sm$^{11}$B$_6$ at 5 K.  (b)   La$^{11}$B$_6$ at 5 K.   (c)  Feynman diagrams illustrating the slave-boson treatment of $f$-electron repulsion within $f$-$d$ hybridized bands as described in the main text. (d)  $Q$dependence of $\chi_0({\bf Q})|F({\bf Q})|^2$, where $\chi_0({\bf Q})$ is the Lindhardt susceptibility for the band structure in (e) and $F(Q)$ is the $5d$ electron form factor.  (e)  Phenomenological band structure within the $(hk0)$ plane.  Translation from $X$ to $M$ shows the change in band character. Inset, schematic representation of pseudonesting vectors.  (f) Smallest unique portion of the Brillouin zone. }
 \end{figure*}

The low energy magnetic neutron scattering cross section for SmB$_6$ is dominated by a resonant mode near 14 meV with bandwidth  $<2$ meV.   Previous publications reported intensity at $R$ $[ (\frac{1}{2}\frac{1}{2}\frac{1}{2})]$, and investigated it versus temperature and doping \cite{oldTAS1, oldTAS2, oldTAS3, oldTAS4, oldTAS5, alekseev1}. Here, we show the mode is also intense near the $X$ $[(\frac{1}{2} 0 0)]$ point and present, albeit dramatically weaker, beyond the first zone.  Through this mulitzone mapping, we provide evidence for an anomalous 5$d$ form factor for the weakly dispersing mode, and develop a minimal band structure based on dominant third neighbor hopping. The hybridized tight-binding model goes beyond early two-band theoretical treatments \cite{Earlytheory1, RpointTheory} by allowing $f$-electron fluctuations as appropriate for a mixed valence compound and provides a link between the wave vector dependence of the magnetic neutron scattering and band inversion in Kondo insulators. Treating $f$-electron Coulomb repulsion with the slave-boson method results in an interaction-protected bound state with dispersion similar to the experiment. 

 SmB$_6$ has Pm3m symmetry with an octahedron of boron in the center of the simple cubic unit cell ($a$ = 4.13 \AA).  Our single crystal was grown by the floating zone method using the non-neutron-absorbing isotopes $^{154}$Sm and $^{11}$B by Yu Paderno and E. Konovalova and initially adopted for lattice \cite{PadernoLattice} and magnetic \cite{alekseev1} inelastic neutron scattering studies on triple-axis spectrometers. We used the SEQUOIA time of flight spectrometer at the SNS with incident energies and elastic energy resolution, respectively, of (50, 2), (80, 2), and (100, 3)~meV \cite{Granroth2006, Granroth2010,  hahn2014inelastic}. Intensity was scaled to absolute units for the differential scattering cross section by normalizing to acoustic phonons and Bragg peaks \cite{funahashi2010x}. 
\begin{figure}
\includegraphics[totalheight=0.245\textheight,viewport= 133 152 770 575,clip]{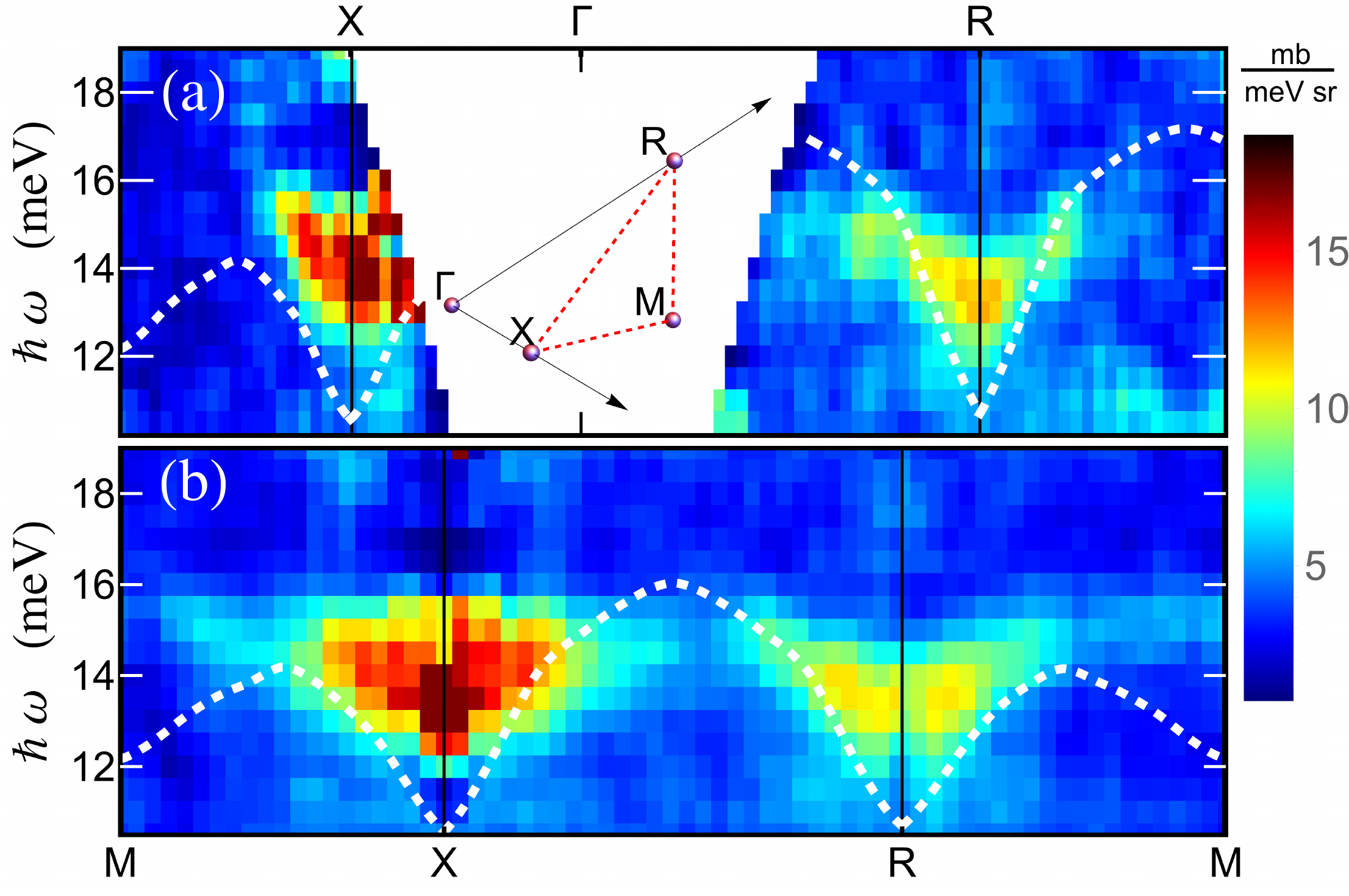}
\caption{ \label{qeSlices}  Neutron scattering cross section for SmB$_6$ at 5 K along high symmetry directions (inset) a) from the $\Gamma$ point and b) around the Brillouin zone edges.  
Dashed line shows the dispersion of a slave-boson-mediated exciton. }
 \end{figure}

Figure~\ref{14 meV} shows the $\bf Q$ dependence of the inelastic scattering intensity, integrated from 12 to 16~meV. Visible at the $R$ point is the intensity maximum previously associated with an intermediate-radius exciton \cite{kikoin1995magnetic}, which reflects the mixed valence state of Sm. The small angle scattering capabilities of SEQUOIA now bring a strong peak at the $X$ point into view, which is replicated at $X$+$G$ $= (\frac{1}{2}1 0)$. The intensity is greatly diminished beyond the first Brillouin zone, indicating the associated spin density extends beyond the 4$f$ orbital (Fig.~\ref{FFplot}).  

\begin{figure}
\includegraphics[totalheight=0.3\textheight,viewport= 5 170 820 827,clip]{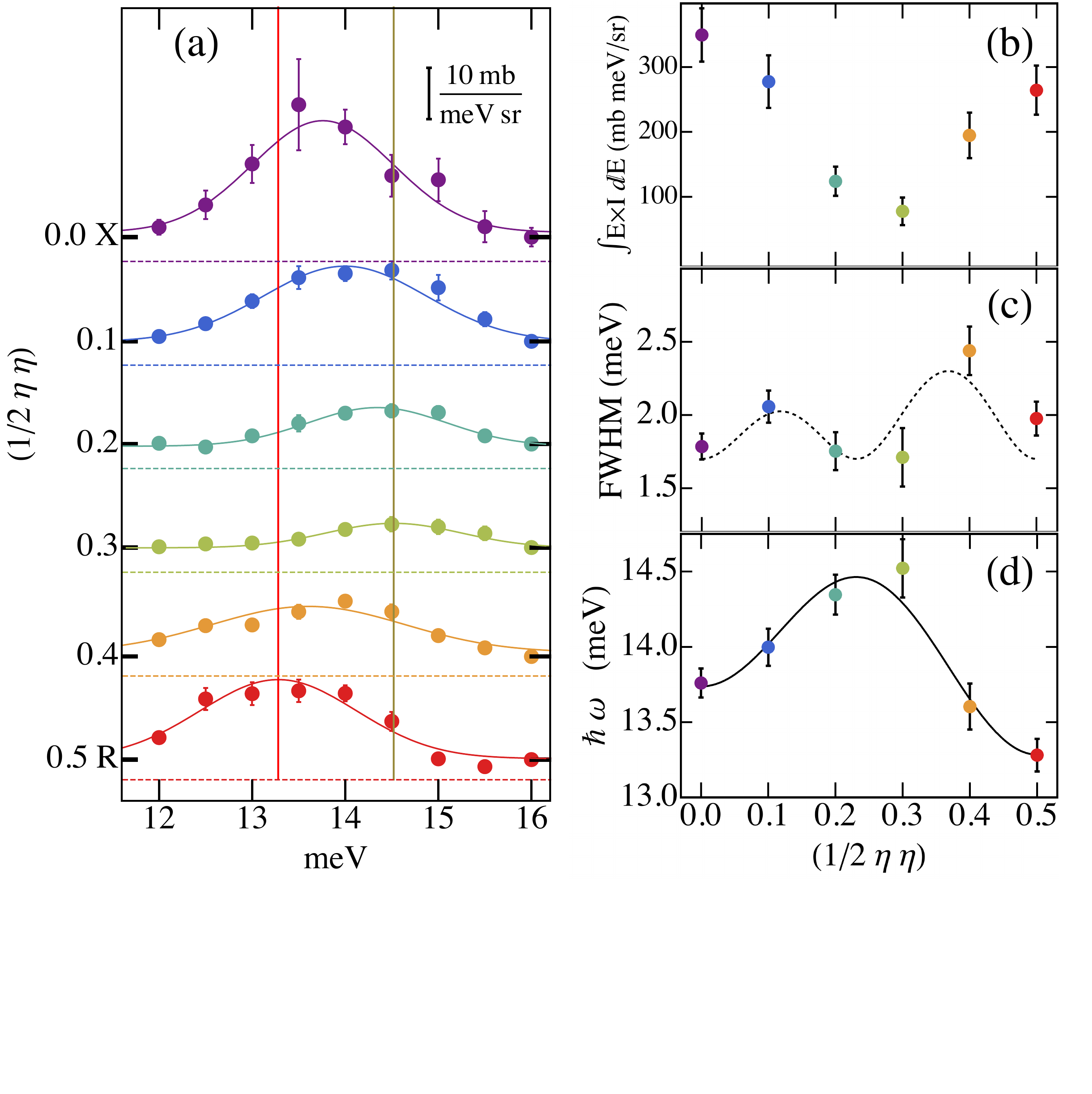}
\caption{ \label{dispersion}  (a) Dispersion along the high-symmetry path between $X$ and $R$. 
Vertical lines show bandwidth.  (b) Oscillator strength, (c) width and resolution(dashed), and (d) peak energies of the 14 meV mode (line is a lattice-sum fit  to guide the eye).}
 \end{figure}

\begin{figure}
\begin{center}
\includegraphics[totalheight=0.29\textheight,viewport= 2 1  290 236,clip] {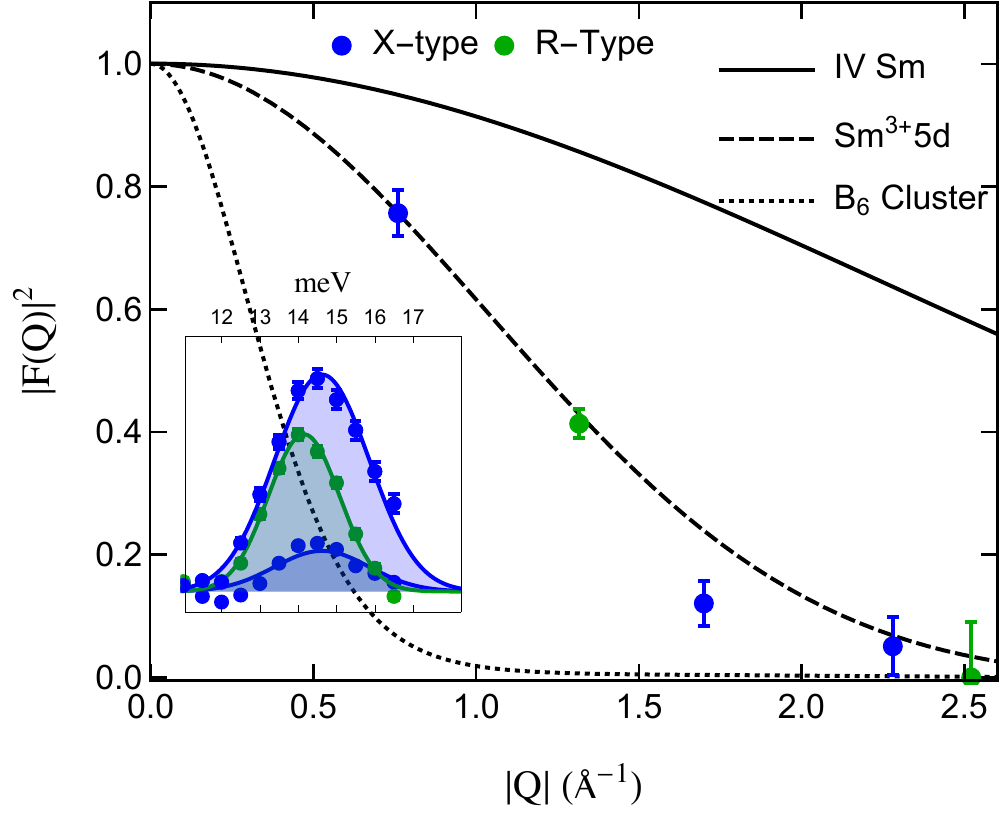}
\caption{ \label{FFplot}  Square of the magnetic form factor of potential scattering centers (lines) and integrated neutron scattering intensity from 12 to 16 meV (symbols) versus $Q$.  Each point is integrated over a cubic $Q$-space volume with side lengths of 0.3 rlu.  Inset: raw data associated with the three smallest $Q$ points. }
\end{center}
 \end{figure}

Figure~\ref{qeSlices} shows the $\bf Q$-dependent spectrum of the neutron scattering intensity along high-symmetry paths though the Brillouin zone. Intensity is confined to regions near the $X$ and $R$ points where the mode energy is minimal. The overall bandwidth of the resonance is less than 2~meV. ~Figure.~\ref{dispersion} provides a quantitative overview of the resonant mode. All peaks in energy transfer are resolution limited [dashed line in (c)], indicating a long-lived collective mode that is isolated from the electron-hole pair continuum. The oscillator strength halfway between $X$ and $R$ falls to less than 20$\%$ of peak values without significant broadening [Fig.~\ref{qeSlices}(b) and Fig.~\ref{dispersion}(b)].  This confinement in momentum space contrasts with a conventional crystal field exciton for which the oscillator strength is $\bf Q$ independent \cite{jensen1991rare}.  

When the magnetic ion forms a simple Bravais lattice, as for $\rm SmB_6$, Bloch's theorem implies $I(\bm q+ \bm G) = I(\bm{q})\times |F(\bm{q}+\bm{G})|^2$, where the form factor $F(\bm{Q}) = {<j_0>}+ (1-\frac{2}{g}) {<j_2>}$, and ${<j_n>} = \int_0^\infty dr^2 r^2 [\rho(r)]^2 j_n(q r)$.  Here $\rho(r)$ is the radial density, ${j_n}$ is the $n$th spherical Bessel function, and $g$ is the Land\'{e} factor \cite{shirane2002neutron}.  We compare the experimental result to the form factors of potential magnetic scattering centers.  Samarium is of mixed valence, with magnetism resulting from the $\rm Sm^{3+} $($J=\frac{5}{2}$) state;  however, the data are inconsistent with the intermediate-valence (IV) form factor that describes the wave vector dependence of field induced magnetic Bragg scattering \cite{SmB6FormFac}. The B$_6$ octahedron would be a magnetic scattering center if the origin of the scattering were electron transfer [Sm$^{2+}$(B$_6$)$^{2-}$ and Sm$^{3+}$(B$_6$)$^{3-}$]; this can be ruled out as the corresponding (B$_6$)$^{3-}$ form factor is indistinguishable from zero beyond the first Brillouin zone, while we observe resonance intensity at $X$+$G$. Instead, the data follow the $5d$ electron form factor (Fig.~\ref{FFplot}) \cite{SMref}, indicating a critical role for such orbitals in the exciton. 

Integrating the exciton scattering over a full Brillouin zone in the energy range from 12 to 16~meV yields the total effective moment: 
$(\mu_{\rm{eff}}/\mu_B)^2 = \int \int \rm{Tr} (S^{\alpha \beta}(\bm Q, \omega)) d^3\bm Q \hbar d\omega/ \int d^3\bm Q =  0.29(6)/$Sm. This corresponds to $\approx40\%$ of the total magnetic scattering cross section for $\rm Sm^{3+}$ \cite{Paderno_1967}. This is a sizable portion of the $>50\%$ of Sm in the 3+ state \cite{PhysRevLett.24.383, mizumaki2009temperature}. The exact valence of Sm in our sample is unknown, but is likely increased due to samarium vaporization during floating-zone growth.

Because the wave vector dependence of the magnetic neutron scattering detected suggests interpretation in terms of a band picture, we proceed to develop a minimal phenomenological model. The nearest electron density to samarium is the B$_6$ cluster.   The lowest energy unoccupied molecular orbitals of nonmagnetic B$_6 ^{2-}$ extend perpendicular to opposing faces of the octahedra in a t$_{1u}$ state. This allows for efficient superexchange along the body diagonal in the magnetic Sm$^{3+}$(B$_6$)$^{3-}$ state. For simplicity we therefore consider a band structure with only third neighbor hopping. 

Although the chemical potential lies in a gap so there is no Fermi surface and no nesting in the conventional sense, $5d$-electron ``pseudonesting" (PN) is expected to enhance the {\it finite energy} generalized susceptibility, and hence to be manifested in the inelastic magnetic neutron scattering through interband transitions.  $X$ and $R$ PN is inherent to a wide range of tight binding band structures dominated by third neighbor hopping. 
 
The $4f$ bands may likewise be assumed to be dominated by third neighbor hopping, albeit with a much smaller bandwidth. To retain a full insulating gap under $f$-$d$ hybridization, the $f$ and $d$ hopping amplitude must have opposite sign. This ensures a gap between the hybridized bands with extrema near the $d$-$f$ band intersections [inset to Fig.~\ref{14 meV}(e)].  The corresponding inter-band transitions now yield PN.  An $X$-type PN boundary is for example visible in Fig.~\ref{14 meV}(e) between regions of hybridization.  

The corresponding phenomenological band structure contains deep band-inversion pockets at $X$ points and a gap of ~15 meV, consistent with ARPES \cite{neupane2013surface}. Because of the fourfold degeneracy of the bands at the $\Gamma$ and $R$ points, only the $X$ and $M$ points contribute to the 3D topological invariant \cite{CubicTKI}, so the proposed phenomenological band structure {\it is} topologically nontrivial. The TKI nature is in fact inherent to a hybridized band structure formed by bands with opposite signs for the dominant third neighbor hopping amplitudes.  

When modulated by the $5d$ electron form factor, the static susceptibility calculated from the resultant particle-hole Green's function is consistent with  the wave vector dependent intensity of the energy integrated inelastic neutron scattering [Fig.~\ref{14 meV}(d)].  Relative scattering strength calculated for the $X$ and $R$ points [Fig.~\ref{14 meV}(d)] is consistent with the neutron data (Fig.~\ref{14 meV}(a)) indicating a similar density of states for both PN wave vectors.  In our third neighbor model, $X$ and $R$ intensity result from PN between cubic faces and as such have nearly identical DOS.  Thus the experimental results in Fig.~\ref{14 meV}(a) and Fig.~\ref{FFplot} support dominant third neighbor hopping. 

From this analysis it is apparent that the wave vector dependence of the inelastic magnetic neutron scattering holds information about band topology. The 14 meV spin exciton we have observed is associated with transitions across the hybridization gap where sharply dispersing $d$ bands define inversion pocket boundaries [Fig.~\ref{14 meV}(e)].  The symmetry of the corresponding patch of enhanced magnetic neutron scattering matches that of its location within the Brillouin zone.  Thus, the observation of magnetic scattering at $X$ with $D_2$ symmetry is associated with $X$-point band inversion, while the absence of intensity along $\Gamma$-$M$ precludes a band inversion at $M$. In cubic TKI only the $X$ and $M$ points contribute to the topological invariant \cite{CubicTKI}. Our analysis of the scattering data thus implies a topologically nontrivial band structure for $\rm SmB_6$.  Comprehensive neutron scattering data combined with such reasoning and comparison to Lindhardt susceptibilities for putative band structures may facilitate analysis of other potential TKIs such as cubic YbB$_6$ and PuB$_6$ \cite{PhysRevLett.111.176404, PhysRevLett.112.016403}, as well as lower symmetry TI candidates.

The collective mode we observed can be understood as an exciton drawn from the electron-hole pair continuum by Coulomb interactions and protected against decay by the hybridization band gap \cite{Riseborough1, Riseborough2}. An exciton forms when the band gap is narrow and the $f$ bandwidth is much smaller than the interactions. The minimal second-quantized Hamiltonian for SmB$_6$ is formulated on the lattice of Sm atoms:
\begin{eqnarray}\label{SmB6Model}
H \!&=&\! \int\limits_{\textrm{1BZ}}\frac{d^{3}k}{(2\pi)^{3}}\biggl\lbrack
     \sum_{\sigma}\xi_{\bf k}^{\phantom{\dagger}}d_{\sigma{\bf k}}^{\dagger}d_{\sigma{\bf k}}^{\phantom{\dagger}}
   +\sum_{\alpha}\epsilon_{\alpha{\bf k}}^{\phantom{\dagger}}f_{\alpha{\bf k}}^{\dagger}f_{\alpha{\bf k}}^{\phantom{\dagger}} \\
&&\!\!\!\!\!\!\!\! +\sum_{\alpha\sigma} \Bigl(V_{\sigma\alpha{\bf k}}^{\phantom{\dagger}}
     d_{\sigma{\bf k}}^{\dagger}f_{\alpha{\bf k}'}^{\phantom{\dagger}}+H.c.\Bigr)\biggr\rbrack
   +U\sum_{\alpha\beta{\bf R}}f_{\alpha{\bf R}}^{\dagger}f_{\alpha{\bf R}}^{\phantom{\dagger}}
     f_{\beta{\bf R}}^{\dagger}f_{\beta{\bf R}}^{\phantom{\dagger}} \nonumber \ .
\end{eqnarray}
Here $d_{\sigma{\bf k}}$ are $d$-electron field operators indexed by spin $\sigma\in\lbrace\uparrow,\downarrow\rbrace$, and $f_{\alpha{\bf k}}$ are $f$-electron operators labeled by the crystal-field multiplet index $\alpha$, which takes into account strong spin-orbit coupling within Sm. Crystal fields introduce $d-f$ hybridization $V$ which produces the narrow band gap. Coulomb repulsion is most influential on the narrow band $f$ electrons, suppressing double occupancy. We thus model interactions by on-site repulsion $U$ among $f$ electrons only. The slave-boson approximation ($U\to\infty$) removes the interaction term in favor of an explicit no-double-occupancy constraint imposed on every site by the auxiliary slave-boson field. The quantum fluctuations of slave bosons renormalize the spectrum and give rise to exciton pairing. These effects can be calculated perturbatively using the random-phase approximation \cite{RPA}.

The perturbation theory is built on top of a mean-field condensate of slave bosons, which shrinks the hybridization band. Slave boson fluctuations introduce further renormalizations of the band structure, which we neglect, and provide the pairing glue for the excitons, which we retain. Figure \ref{14 meV}(c) shows the associated Feynman diagrams:  diagrams 1 and 2 show the $f$-$d$ hybridization process wherein the slave-boson-mediated conversion between an $f$ and a $d$ electron [diagram 1] dominates over $f$ electron scattering on slave bosons [diagram 2]. This resonant conversion provides electron-hole pairing glue that stabilizes an exciton as illustrated in diagram 3. Self-energy corrections [diagram 4] shrink the exciton bandwidth and produce the relatively flat collective mode seen in the experiment. The self-consistently renormalized slave-boson propagator in diagram 4 stands for all wavy lines in diagram 3; its numerical properties are extracted from experimental data by a fitting procedure described in detail elsewhere \cite{nikolic2014two, fuhrman2014gap, inPrep}. 

Figure~\ref{qeSlices} compares our experimental results with the calculated exciton dispersion. Since the precise microscopic values of parameters are not known, we fit their renormalized values to match the calculated and measured spectra. Using the band structure described above, the calculated exciton dispersion relation is consistent with the experiment, having comparable bandwidth and minima at high symmetry points. The existence of the exciton and its apparent origin in Coulomb interactions portray $\rm SmB_6$ as a correlated (Mott) insulator where the lowest energy excitations are bosonic rather than fermionic as in band insulators.  

We observed a 14 meV collective mode in an extensive region of momentum space that we describe as a slave-boson-mediated bound state.  The $5d$-like exciton form factor is evidence for a significant role of $5d$ orbitals in the exciton while the symmetry of the high intensity regions in momentum space reflects a topologically nontrivial renormalized band structure consistent with higher energy ARPES data. This exciton is a consequence of the protection afforded by correlations within an insulator born of hybridization; the Kondo singlet fluctuations it represents show that correlations drive the TI phase in SmB$_6$ and as a consequence we can expect the toplogically protected surface states of SmB$_6$ to exhibit strongly correlated 2D physics. 
 
This project was supported by UT-Battelle LDRD \#3211-2440.  The work at IQM was supported by the U.S. Department of Energy, Office of Basic Energy Sciences, Division of Material Sciences and Engineering under Grant No. DE-FG02-08ER46544. Research conducted at ORNL's SNS was sponsored by the Scientific User Facilities Division, Office of Basic Energy Sciences, U.S. Department of Energy.  P.A.A. is grateful to RFBR Grant No.14-22-01002 for the partial support of participation in this work.  The authors thank Martin Mourigal, Yuan Wan, and Ari Turner for fruitful discussions.  

\normalfont\textsuperscript{$\star$} W.T.F. and J.L. contributed equally to this work.  

\bibliography{SmB6_ref}

\newpage
\onecolumngrid
\appendix
\beginsupplement
\section{{SUPPLEMENTARY MATERIAL}}

\subsection{Resolution Volume}
In order to assess the possible loss of intensity in Fig. 4 due to mosaic distribution in the crystal, we examined the consistency of the Bragg intensities with respect to their structure factors.  Fig.~\ref{ResVol} below shows the wave vector dependence of nuclear Bragg intensities normalized to the corresponding squared structure factor following identical symmetrization and integration procedures as that used for the magnetic scattering. The $Q-$independence of these data shows the experiment and analysis procedures accurately probe the cross section despite the effects of absorption and crystal mosaic. 
\begin{figure}[H]
\begin{center}
\includegraphics[totalheight=0.25\textheight,viewport= 0 0 359 190,clip] {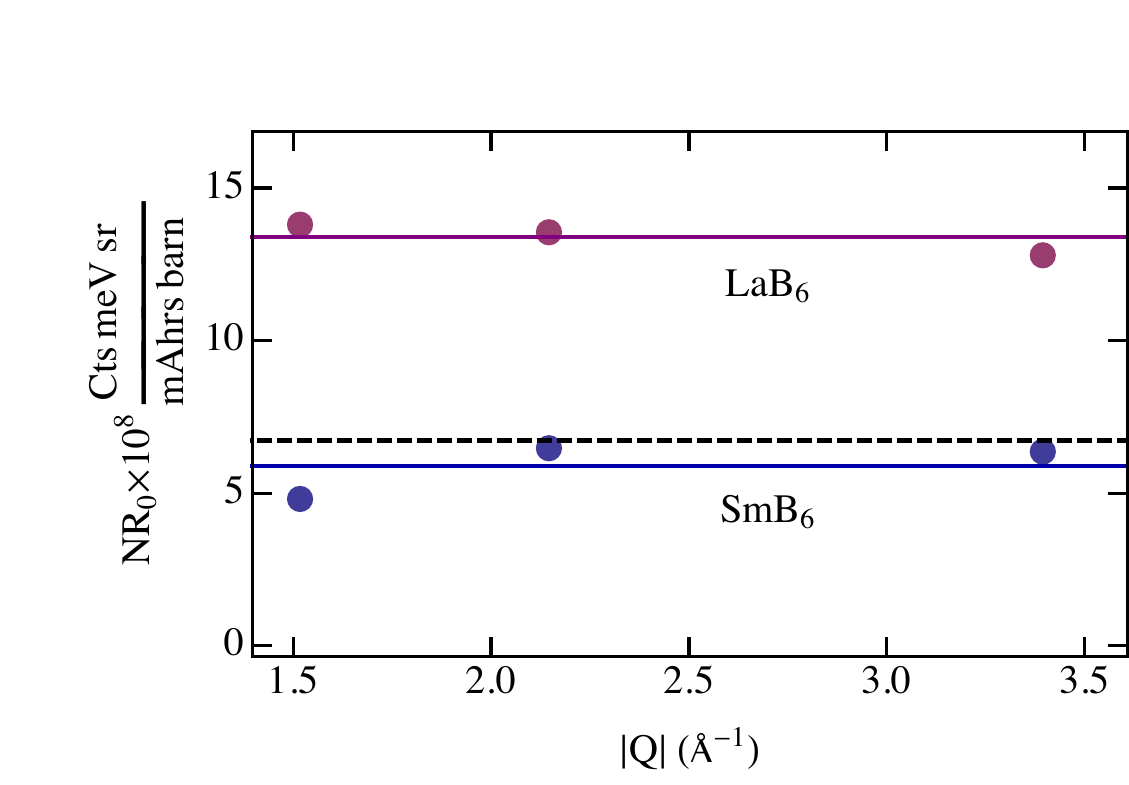}
\caption{ \label{ResVol}  Resolution volume inferred from the integrated Bragg diffraction intensity. To account for different sample masses, the $\rm LaB_6$ data was scaled by a factor of .5.  The smaller resolution volume for SmB$_6$ is primarily due to residual content of neutron absorbing samarium isotopes. Solid line is average, dashed line shows the value derived from the intensity of one-phonon neutron scattering for SmB$_6$.}
\end{center}
 \end{figure}

\end{document}